\documentclass[twocolumn,amsmath,amssymb]{revtex4}
\usepackage{graphicx}% Include figure files
\usepackage{dcolumn}% Align table columns on decimal point
\usepackage{bm}% bold math

\begin{document}

%\preprint{APS/123-QED}

\title{Counting statistics for genetic switches based on effective interaction approximation}
\author{Jun Ohkubo}
\email[Email address: ]{ohkubo@i.kyoto-u.ac.jp}
\affiliation{
Graduate School of Informatics, Kyoto University,\\
Yoshida Hon-machi, Sakyo-ku, Kyoto-shi, Kyoto 606-8501, Japan
}
%\date{\today}% It is always \today, today,
             %  but any date may be explicitly specified

\begin{abstract}
Applicability of counting statistics for a system with an infinite number of states is investigated.
The counting statistics has been studied a lot for a system with a finite number of states.
While it is possible to use the scheme in order to count specific transitions
in a system with an infinite number of states in principle, 
we have non-closed equations in general.
A simple genetic switch can be described by a master equation 
with an infinite number of states,
and we use the counting statistics in order to 
count the number of transitions from inactive to active states in the gene.
To avoid to have the non-closed equations,
an effective interaction approximation is employed.
As a result, it is shown that the switching problem can be treated as a simple two-state model approximately,
which immediately indicates that the switching obeys non-Poisson statistics.
\end{abstract}

%\pacs{}% PACS, the Physics and Astronomy
                             % Classification Scheme.
%\keywords{Suggested keywords}%Use showkeys class option if keyword
                              %display desired
\maketitle

\section{Introduction}
\label{sec_introduction}

Counting statistics is a scheme to calculate all statistics related to specific transitions
in a stochastic system.
In the counting statistics, a master equation with discrete states is used 
to derive time-evolution equations for generating functions related to the specific transitions.
The scheme has been used to investigate F{\"o}rster resonance energy transfer,
and many successful results have been obtained \cite{Gopich2003,Gopich2005,Gopich2006}.
Although the scheme is basically formulated for a system with a finite number of states,
it is possible to use the scheme to investigate a system with an infinite number of states.
However, as exemplified later, we have non-closed equations in general,
so that it would be needed to develop approximation schemes suitable for specific systems.
As a first step, it is important to check 
whether an approximation scheme for the counting statistics is available for
the system with an infinite number of states or not.

In the present paper, we focus on dynamics in genetic switches.
It has been shown that stochastic behavior 
plays an important role in gene regulatory systems \cite{Elowitz2002,Rao2002,Kaern2005},
and there are many studies
for the stochasticity in the gene regulatory systems
from experimental points of view (e.g., see \cite{Gardner2000,Okano2008})
and theoretical ones (e.g., see  
\cite{Hasty2000,Sasai2003,Hornos2005,Xu2006,Schultz2007,Shahrezaei2008,Walczak2009,Venegas-Ortiz2011}).
Not only studies by numerical simulations,
but also those by analytical calculations have been performed.
Some analytical expressions for 
the static properties, i.e., stationary distributions
for the number of proteins or mRNAs, have already been obtained.
In addition, in order to investigate the role of the stochasticity in genetic switches,
dynamical properties, i.e., switching behavior between active and inactive gene states,
have also been studied.
Basically, such dynamical properties have been investigated 
by numerical simulations (e.g., see \cite{Feng2011});
only for a simple system, analytical expressions for the first-passage time distribution
have been obtained \cite{Visco2009}.
The genetic switch is described by a master equation with an infinite number of states.
Hence, if we can use the scheme of the counting statistics
in order to investigate the dynamical properties in the genetic switches,
it will be helpful to obtain deeper understanding and intuitive pictures
for the genetic switches.

The aim of the present paper is 
to seek the applicability of the counting statistics
in order to investigate the dynamical property in the genetic switches.
It immediately becomes clear that
a straightforward application of the counting statistics derives intractable non-closed equations.
In order to obtain simple closed forms,
we here employ an effective interaction approximation \cite{Ohkubo2011}.
As a result, we will show that the switching problem
can be treated as a simple two-state model approximately.
This result immediately gives us intuitive understanding for the switching behavior
and the non-Poissonian property.
% In addition, this simple approximate description would enable us 
% to treat the switching problems for more complicated gene regulatory systems.

The present paper is constructed as follows.
In Sec.~\ref{sec_model}, we give a brief explanation of a stochastic model for the genetic switch.
In Sec.~\ref{sec_counting_statistics}, 
the counting statistics is employed in order to count
the number of transitions in the genetic switch,
and, as a result, a simple two-state model is derived approximately.
The derived approximated results
are compared with those of Monte Carlo simulations in Sec.~\ref{sec_results}.
Section \ref{sec_conclusions} gives concluding remarks.

\section{Model}
\label{sec_model}

\begin{figure}
\begin{center}
\includegraphics[width=60mm]{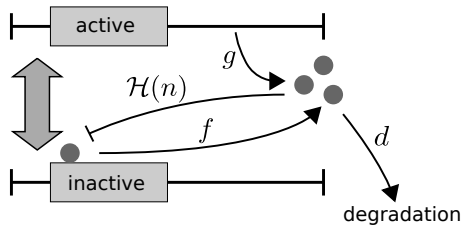}
\end{center}
\caption{
A schematic illustration of the self-regulating gene with repressed binding interaction.
When the regulatory proteins are combining the gene, 
the gene is repressed and there is no production of proteins.
If the regulatory proteins are released from the gene,
the gene becomes active and it can produce the proteins.
We consider the transition between the active and inactive states
as a switch.
}
\label{fig_model}
\end{figure}

A gene regulatory system consists of many components, such as genes, RNAs, and proteins.
Here, a simplified model is used; mRNAs are neglected for simplicity,
and an activated gene assumes to directly increase the number of proteins.
In addition, in the simplified model, a repressed gene cannot produce any proteins.
The above model has been used to investigate the switching behavior
in previous works,
and, for example, see \cite{Schultz2007} for details of the model.

We summarize the model studied in the present paper in Fig.~\ref{fig_model}.
The binding interaction is assumed to be a repressed one,
and the gene is activated only when the regulatory proteins are not binding the gene.
The proteins are produced from the gene in the active state with rate $g$,
and proteins are degraded spontaneously with rate $d$.
The regulatory proteins bind the gene with a rate function $\mathcal{H}(n)$,
where $n$ is the number of free proteins.
For example, $\mathcal{H}(n) = h n$ for a monomer interaction case,
and $\mathcal{H}(n) = h n(n-1)/2$ for a dimer interaction case,
where $h$ is a rate constant for the binding.
$f$ is a rate constant with which the regulatory proteins are released
from the repressor site of the gene.

% start
We here give short comments for the model from the viewpoint of experiments.
Using this simplified model,
we can discuss the connection among the model parameters,
the number of proteins, and the switching behaviors.
While the number of proteins $n$ can be observed or estimated experimentally,
as far as we know, there has not been an experimental technique
to observe the attachment and detachment of the regulatory proteins directly.
We hope that developments of single-molecule observations in future
would enable us to give information about the switching dynamics.
% end

\section{Counting statistics for the number of transitions}
\label{sec_counting_statistics}

\subsection{Master equation for the number of proteins}
\label{subsec_master_equation}

Analytical treatments for the self-regulating gene system have been developed,
and an exact solution is known for the monomer interaction case, i.e., 
$\mathcal{H}(n) = h n$ \cite{Hornos2005,Visco2009}.
In order to simplify the analytical treatments,
an additional assumption has been used in some previous works \cite{Schultz2007,Ohkubo2011};
i.e., some of proteins are assumed to be inert when the gene state is active.
The inert proteins cannot repress the gene, and it is not degraded.
For the monomer interaction case, there is only one inert protein;
the number of inert protein for the dimer interaction case is two, and so on.
% added start
Note that the assumption of the inert proteins does not have physical meanings;
this only simplify the analytical treatments (for details, see \cite{Schultz2007}).
However, 
% added end
it has been shown that this assumption has little influence of the gene system,
and then we employ the assumption in the present paper.

Let $\alpha_n$ and $\beta_n$ be states 
in which there are $n$ free proteins for the active and inactive states, respectively.
The probabilities for $\alpha_n$ and $\beta_n$ at time $t$ satisfy
the following master equations;
\begin{align}
\frac{d P(\alpha_n,t)}{dt} =
& g [ P(\alpha_{n-1},t) - P(\alpha_n,t) ] \nonumber \\
& + d [(n+1) P(\alpha_{n+1},t) - n P(\alpha_n,t) ] \nonumber \\
&- h n P(\alpha_n,t) + f P(\beta_n,t), 
\label{eq_master_monomer_1_exact}\\
\frac{d P(\beta_n,t)}{dt} =
& d [(n+1)P(\beta_{n+1},t) - n P(\beta_n,t)]  \nonumber \\
&+ h n P(\alpha_n,t) - f P(\beta_n,t),
\label{eq_master_monomer_2_exact}
\end{align}
where $P(\alpha_n,t)$ and $P(\beta_n,t)$
are probabilities for $n$ free proteins for the active and inactive states,
respectively.

As stated in Sec.~\ref{sec_introduction},
the exact solutions for stationary distributions
of the number of proteins have been derived,
and those are expressed
using the Kummer confluent hypergeometric functions.
For details, see \cite{Hornos2005,Schultz2007}.

\subsection{Counting statistics}
\label{subsec_counting_statistics}

Using the concept of the counting statistics \cite{Gopich2003,Gopich2005,Gopich2006},
it is possible to investigate dynamical properties, i.e.,
all statistics for the switching behavior between the active and inactive states.
In the present paper, as an example, we calculate
the number of transitions from the inactive state to the active state.
The generating functions for the transitions
are immediately obtained from the master equations \eqref{eq_master_monomer_1_exact}
and \eqref{eq_master_monomer_2_exact}.
A brief explanation of the counting statistics is given in the Appendix,
and we here give consequences of the counting statistics.

A probability, with which there are $k$ transitions from the inactive state to the active state
during time $t$, is denoted by $P(k|t)$.
The generating function for $P(k|t)$ is defined as
\begin{align}
F(\lambda,t) = \sum_{k=0}^{\infty} P(k|t) \lambda^{k},
\end{align}
where $\lambda$ is a counting variable.
The generating function gives all information related to ``inactive $\to$ active'' transitions.
According to the scheme of counting statistics,
we split $F(\lambda,t)$ into restricted generating functions 
$\{\phi(\alpha_n,\lambda,t)\}$ and $\{\phi(\beta_n,\lambda,t)\}$,
where $\phi(\alpha_n,\lambda,t)$ and $\phi(\beta_n,\lambda,t)$ are 
the generating functions for the system
in states $\alpha_n$ and $\beta_n$ at time $t$, respectively.
Using the scheme of the counting statistics,
we obtain the following time-evolution equations for the restricted generating functions
$\{\phi(\alpha_n,\lambda,t)\}$ and $\{\phi(\beta_n,\lambda,t)\}$:
\begin{align}
\frac{d \phi(\alpha_n,\lambda,t)}{dt} =
& g [ \phi(\alpha_{n-1},\lambda,t) - \phi(\alpha_n,\lambda,t) ] \nonumber \\
& + d [(n+1) \phi(\alpha_{n+1},\lambda,t) - n \phi(\alpha_n,\lambda,t) ] \nonumber \\
&- h n \phi(\alpha_n,\lambda,t) + \lambda f \phi(\beta_n,\lambda,t), 
\label{eq_cs_1_exact}\\
\frac{d \phi(\beta_n,\lambda,t)}{dt} =
& d [(n+1)\phi(\beta_{n+1},\lambda,t) - n \phi(\beta_n,\lambda,t)]  \nonumber \\
&+  h n \phi(\alpha_n,\lambda,t) - f \phi(\beta_n,\lambda,t).
\label{eq_cs_2_exact}
\end{align}
Although Eqs.~\eqref{eq_cs_1_exact} and \eqref{eq_cs_2_exact}
are similar to Eqs.~\eqref{eq_master_monomer_1_exact} and \eqref{eq_master_monomer_2_exact},
note that the final term in the right hand side of Eq.~\eqref{eq_cs_1_exact} has
a factor $\lambda$.
% add start
The factor $\lambda$ is introduced in order to count the number of transitions,
and we can count the number of transitions related to this term (for details, see Appendix).
% end
Using the above restricted generating functions,
the generating function $F(\lambda,t)$ is calculated as
\begin{align}
F(\lambda,t) = \sum_{n=0}^{\infty} 
\left\{ \phi(\alpha_n,\lambda,t) + \phi(\beta_n,\lambda,t) \right\}.
\label{eq_total_generating_function}
\end{align}

%Note that in previous works \cite{Gopich2005,Gopich2006},
%the analytical scheme is applied for only cases with finite number of states.
%In contrast, the number of states in our case is infinite 
%(the number of free proteins $n$ takes an integer value from $0$ to $\infty$).

Next, we introduce the following generating functions 
for $\phi(\alpha_n,\lambda,t)$ and $\phi(\beta_n,\lambda,t)$:
\begin{align}
&\alpha(\lambda,z,t) \equiv \sum_{n=0}^{\infty} \phi(\alpha_n,\lambda,t) z^n,\\
&\beta(\lambda,z,t) \equiv \sum_{n=0}^{\infty} \phi(\beta_n,\lambda,t) z^n.
\end{align}
It is straightforward to derive the time-evolution equations 
for the new generating functions $\alpha(\lambda,z,t)$ and $\beta(\lambda,z,t)$
from Eqs.~\eqref{eq_cs_1_exact} and ~\eqref{eq_cs_2_exact};
\begin{align}
\frac{d \alpha(\lambda,z,t)}{dt} =
& (z-1) \left[ g \alpha(\lambda,z,t) 
- d \frac{\partial \alpha(\lambda,z,t)}{\partial z} \right] \nonumber
\label{eq_cs_1_exact_modified} \\
& - hz \frac{\partial \alpha(\lambda,z,t)}{\partial z} + \lambda f \beta(\lambda,z,t), \\
\frac{d \beta(\lambda,z,t)}{dt} =
& - (z-1) d \frac{\partial \beta(\lambda,z,t)}{\partial z} \nonumber \\
& +  hz \frac{\partial \alpha(\lambda,z,t)}{\partial z} 
- f \beta(\lambda,z,t). \label{eq_cs_2_exact_modified}
\end{align}
Using the generating function $\alpha(\lambda,z,t)$ and $\beta(\lambda,z,t)$,
 the generating function $F(\lambda,t)$ is given by
\begin{align}
F(\lambda,t) = \alpha(\lambda,z=1,t) + \beta(\lambda,z=1,t),
\end{align}
and therefore it is enough to solve the following time-evolution equations
in order to calculate the generating function $F(\lambda,t)$:
\begin{align}
\frac{d \alpha(\lambda,t)}{dt} =&
- h \left. \frac{\partial \alpha(\lambda,z,t)}{\partial z} \right|_{z=1} 
+ \lambda f \beta(\lambda,t), \label{eq_cs_1_reduced_exact}\\
\frac{d \beta(\lambda,t)}{dt} =&
 h \left. \frac{\partial \alpha(\lambda,z,t)}{\partial z} \right|_{z=1} 
- f \beta(\lambda,t), \label{eq_cs_2_reduced_exact}
\end{align}
where we define $\alpha(\lambda,t) \equiv \alpha(\lambda,z=1,t)$
and $\beta(\lambda,t) \equiv \beta(\lambda,z=1,t)$.

Note that Eqs.~\eqref{eq_cs_1_reduced_exact} and \eqref{eq_cs_2_reduced_exact}
contain the derivative of $\alpha(\lambda,z,t)$ with respect to $z$.
Hence, the equations are not closed.
If these terms are expressed simply using $\alpha(\lambda,t)$,
we will have simultaneous differential equations written only by
the generating functions $\alpha(\lambda,t)$ and $\beta(\lambda,t)$;
i.e., we have closed equations and hence
the obtained equations may be solved analytically.
In the following analysis,
an effective interaction approximation is employed,
and we will show that the above statistics
can be approximated by a simple two-state model.

\subsection{Approximation for the interaction}
\label{subsec_approximation}

In the effective interaction approximation,
the interaction function $\mathcal{H}(n)$ is replaced as a constant value.
As shown in \cite{Ohkubo2011}, the dependence of $\mathcal{H}(n)$ on $n$
makes it difficult to obtain analytical results,
and it has been shown that the approximation gives qualitatively good results.

Replacing the interaction function $\mathcal{H}(n)$ as
\begin{align}
\mathcal{H}(n) = \tilde{h},
\end{align}
where $\tilde{h}$ is a constant,
we obtain the following equations instead of 
Eqs.~\eqref{eq_cs_1_reduced_exact} and \eqref{eq_cs_2_reduced_exact}:
\begin{align}
\frac{d \alpha(\lambda,t)}{dt} =&
- \tilde{h} \alpha(\lambda,t) + \lambda f \beta(\lambda,t), 
\label{eq_cs_1_reduced_approximate}\\
\frac{d \beta(\lambda,t)}{dt} =&
 \tilde{h} \alpha(\lambda,t) - f \beta(\lambda,t).
\label{eq_cs_2_reduced_approximate}
\end{align}
Note that Eqs.~\eqref{eq_cs_1_reduced_approximate} and \eqref{eq_cs_2_reduced_approximate}
are written only by $\alpha(\lambda,t)$ and $\beta(\lambda,t)$.
It means that the switching problem can be approximated as
a simple two-state model
\textit{if} the effective interaction $\tilde{h}$ is chosen adequately.

We here briefly explain the choice of the effective interaction $\tilde{h}$
using a simple example, i.e., the monomer binding interaction case.
For the monomer binding interaction,
the interaction function is calculated as follows \cite{Ohkubo2011}.
In this case, the interaction function is $hn$.
In order to obtain the effective interaction $\tilde{h}$,
the number of proteins $n$ is replaced 
as the average number of proteins, i.e., 
\begin{align}
\tilde{h} = h \langle n \rangle_{\alpha},
\label{eq_effective_int_monomer}
\end{align}
where $\langle n \rangle_{\alpha}$ is the expectation of the number of free regulatory proteins
under a condition that the gene is in the active state (conditional expectation).

The conditional expectation can be calculated from 
the stationary distribution of the number of proteins.
Note that the generating functions 
$\alpha(\lambda,z,t)$ and $\beta(\lambda,z,t)$
are reduced to generating functions for the stationary distribution
of the number of proteins when $\lambda = 1$.
Hence, as shown in \cite{Ohkubo2011},
they are written as follows.
\begin{align}
\alpha(z) \equiv& \lim_{t\to \infty}\alpha(\lambda=1,z,t) = A F[a,b,N(z-1)], \\
\beta(z) \equiv& \lim_{t\to \infty}\beta(\lambda=1,z,t) \nonumber \\
= &
\left( 1+ \frac{\tilde{h}}{f} \right) A F[a-1,b-1,N(z-1)] - \alpha(z)
% \nonumber \\
%& - \lim_{t \to \infty}\alpha(\lambda=1,z,t),
\end{align}
where $A = f/(f+\tilde{h})$ and
\begin{align*}
N = \frac{g}{d}, \quad a = 1 + \frac{f}{d}, \quad b = 1 + \frac{f+\tilde{h}}{d}.
\end{align*}
$F(p,q,r)$ is the Kummer confluent hypergeometric function,
\begin{align}
F(p,q,r) \equiv \sum_{n=0}^{\infty} \frac{(p)_n}{(q)_n} \frac{r^n}{n!},
\end{align}
where $(p)_n = p(p+1)(p+2)\cdots (p+n-1)$.
We, therefore, obtain
\begin{align}
\langle n \rangle_{\alpha}
\equiv \frac{1}{\alpha(1)} 
\left. \frac{\partial}{\partial z} \alpha(z) \right|_{z=1}
= \frac{g(d+f)}{d(d+f+\tilde{h})}.
\label{eq_for_effective}
\end{align}
By inserting Eq.~\eqref{eq_for_effective} into Eq.~\eqref{eq_effective_int_monomer},
the following self-consistent equation is derived:
\begin{align}
\tilde{h}
= h \frac{g(d+f)}{d(d+f+ \tilde{h})}.
\label{eq_monomer_self_consistent}
\end{align}
Solving Eq.~\eqref{eq_monomer_self_consistent},
we obtain 
\begin{align}
\tilde{h} = 
\frac{-(d^2 + fd) + \sqrt{(d^2+fd)^2 + 3 h g d(d+f)}}{2d}.
\end{align}

We finally comment on a solution of the simple two-state model 
(Eqs.~\eqref{eq_cs_1_reduced_approximate} and \eqref{eq_cs_2_reduced_approximate}).
The simple two-state model can be solved exactly \cite{Gopich2003,Gopich2006},
and the probability distribution $P(k|t)$ for the number of ``inactive $\to$ active'' transitions
during time $t$ is explicitly written as follows:
\begin{align}
P(k|t) =& \left( \frac{(1-\gamma^2) T}{2\gamma}\right)^k \frac{e^{-T}}{k! \sqrt{8 \gamma T / \pi}} \nonumber\\
&\times \left\{ 2 \gamma (k+T) I_{k-1/2}(\gamma T) 
 + (1+\gamma^2) T I_{k+1/2}(\gamma T )  \right\},
\label{eq_prob_distribution}
\end{align}
where $T = (f + \tilde{h}) t /2$, $\gamma^2 = 1 - 4 f \beta(1)/(f+\tilde{h})$,
and $I_n(z)$ are modified Bessel functions of the first kind.
This expression~\eqref{eq_prob_distribution} immediately gives us
the non-Poissonian picture of the phenomenon.

\section{Numerical results}
\label{sec_results}

\begin{figure}
\begin{center}
\includegraphics[width=70mm]{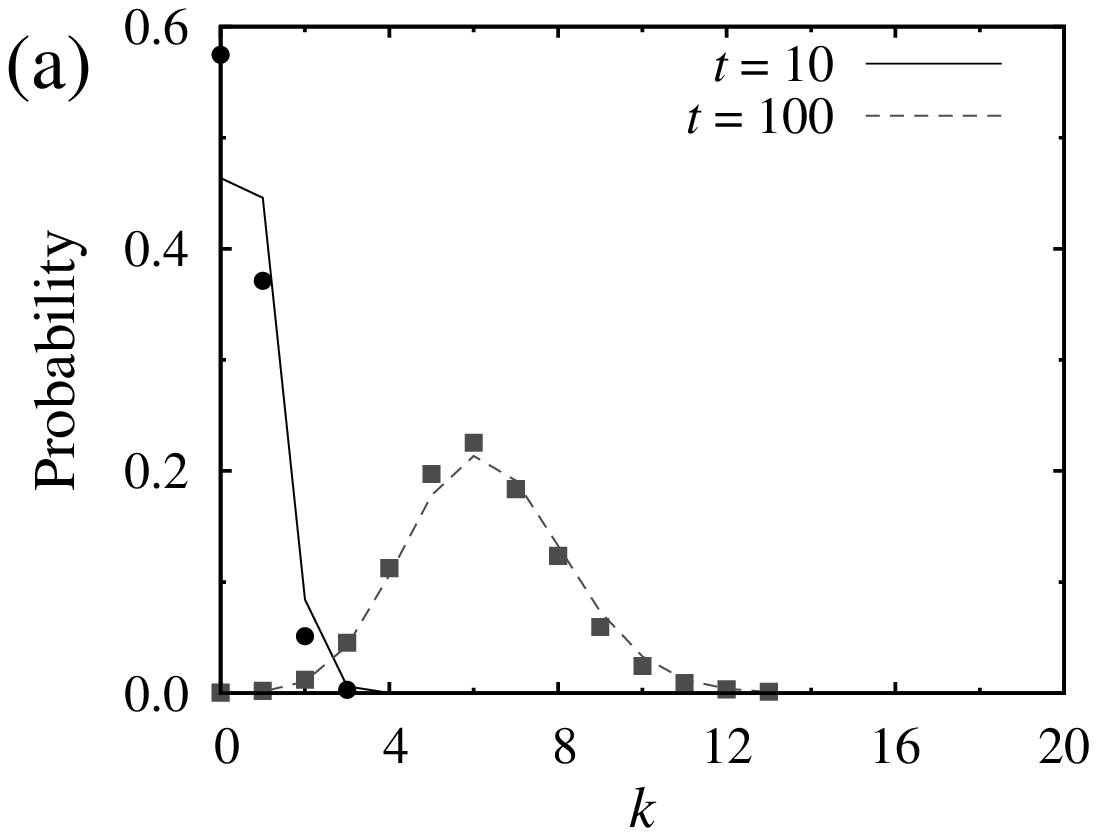}\\
\includegraphics[width=70mm]{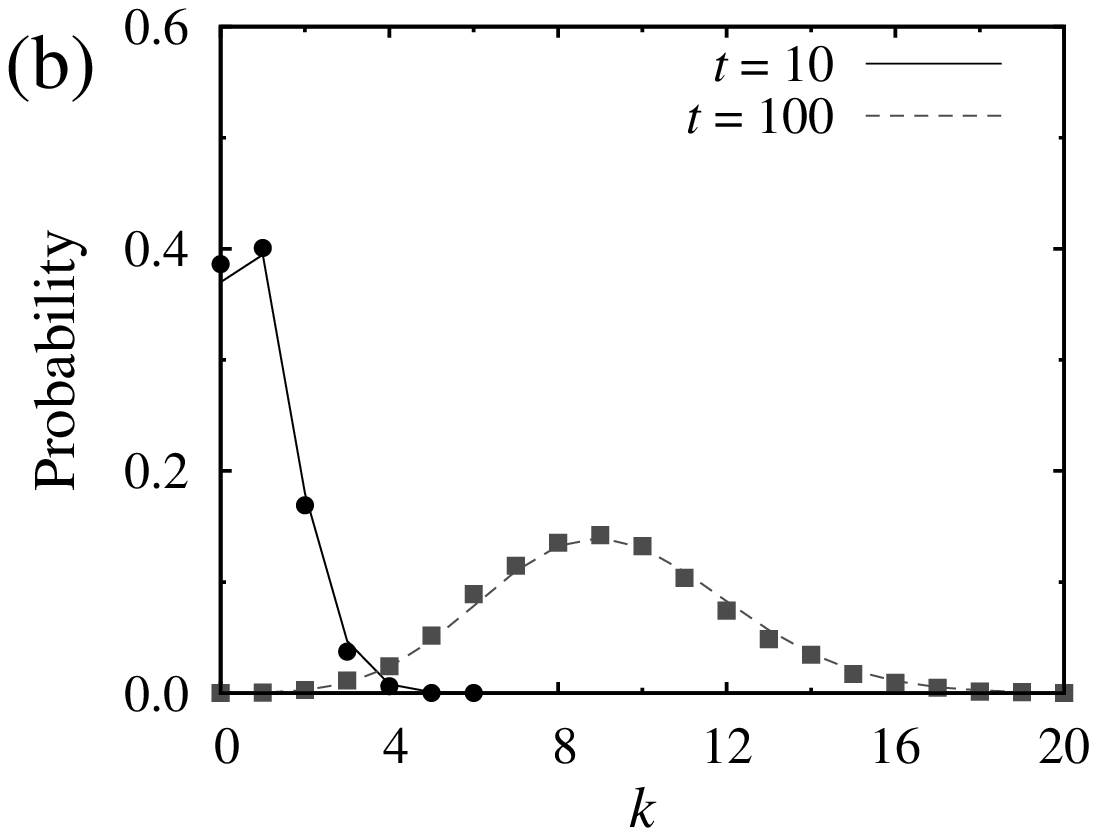}
\end{center}
\caption{Probability distributions for the number of ``inactive $\to$ active'' transitions.
(a) Monomer binding interaction case. (b) Dimer binding interaction case. 
In each figure, filled circles and filled boxes are Monte Carlo results
for time $t = 10$ and $t = 100$, respectively.
Solid and dashed lines corresponds to 
approximated analytical results of Eq.~\eqref{eq_prob_distribution}
for time $t = 10$ and $t = 100$, respectively.
}
\label{fig_results}
\end{figure}

In order to check the validity of the analytical treatments
and the approximations,
we here compare the analytical results with
those of Monte Carlo simulations.
The original genetic switch explained in Sec.~\ref{sec_model}
was simulated using a standard Gillespie algorithm \cite{Gillespie1977}.
The parameters used in the simulation are as follows:
$d = 1, \quad g = 50.0, \quad h = 0.004, \quad f = 0.1$.
Note that these parameters were selected as one of the typical values used 
in the previous works \cite{Schultz2007,Ohkubo2011}.

Firstly, we consider the monomer binding interaction case.
According to the discussions in Sec.~3.3, 
the value of the effective interaction $\tilde{h}$ is calculated as $\tilde{h} = 0.173$.
Figure~\ref{fig_results}(a) shows 
the results of the analytical calculations (Eq.~\eqref{eq_prob_distribution})
and those of the Monte Carlo simulations.
Although there are quantitative differences,
the results shows that the approximated two-state model
captures the essential features of the phenomenon.

Next, we consider a dimer binding interaction case,
i.e., $\mathcal{H}(n) = h n(n-1)/2$.
In this case, the effective interaction is calculated as follows:
\begin{align}
\tilde{h} = h \frac{\langle n (n-1) \rangle_{\alpha}}{2}.
\end{align}
As shown in \cite{Ohkubo2011}, 
the effective interaction $\tilde{h}$ is obtained
by solving the following self-consistent equation:
\begin{align}
\tilde{h}
= \frac{h}{2} \frac{1}{\alpha(1)} 
\left. \frac{\partial^2}{\partial z^2} \alpha(z) \right|_{z=0}.
\label{eq_dimer_self_consistent}
\end{align}
We here numerically solved the self-consistent equation
(Eq.~\eqref{eq_dimer_self_consistent}),
and the calculated value of the effective interaction is $\tilde{h} = 1.358$.
Using the calculated value,
we depict the analytical results and the corresponding Monte Carlo results
in Fig.~\ref{fig_results}(b).
From the comparison,
we confirmed that
the approximated two-state model is available even in the dimer binding interaction case.
% start
Although results are not shown,
we performed numerical simulations for other some parameters,
and checked the validity of the analytical treatments.
For example, even for parameter regions in which 
the probability distribution of the number of proteins has bistability,
the approximation scheme works well.
%Even when the switching occurs rapidly (we have checked numerical results up until $h \sim 1000.0$), 
%Even when the switching occurs rapidly (for example, for a case with $h = 10^3$)
%this approximation works well.
% end

\section{Conclusions}
\label{sec_conclusions}

In the present paper, we studied an analytical scheme 
to extract information related to the dynamical behavior in genetic switches.
Using an effective interaction approximation,
a simple two-state model is obtained,
and we confirmed that the two-state model captures the features of the phenomenon.
Note that in the analytical treatments,
we did not neglect the stochastic properties of the system
(except for the effective interaction approximation);
i.e., we can calculate all statistics for transitions approximately, including higher order moments.
It could be possible 
to apply the above effective expression for the transitions between the active and inactive states
to more complicated gene regulatory networks
without loss of the stochasticity;
this would give us deeper understanding
for the switching behavior of the gene regulatory systems
including static, dynamical, and stochastic behaviors.
% start
In addition, the idea of the effective interaction may be similar
to the mean-field approximation in statistical physics;
the interaction is replaced with the average.
It may be possible to develop higher-order approximations
using the analogy with the conventional approximation schemes in statistical physics;
this is an important future work.
% end

We discussed properties only in the stationary states,
because the effective interaction approximation
has been applied only for the stationary states at the moment;
the average number of proteins (or higher moments) should be estimated adequately,
and it was calculated by using the analytical solutions
for the \textit{stationary} distributions of the number of proteins.
Recently, exact time-dependent solutions for a self-regulating gene have been derived
\cite{Ramos2011}.
Hence, it may be possible to extend the effective interaction approximation
to non-stationary states.
If so, the effective interaction $\tilde{h}$ would be time-dependent,
and, at least numerically, 
it is possible to calculate various moments for the counting statistics
for time-dependent systems \cite{Ohkubo2010}.
We expect that the simple description developed in the present paper
is available for various cases,
such as complicated regulatory systems and time-dependent systems,
and that the description gives new insights
for the regulation mechanisms and stochastic behaviors.

\section*{ACKNOWLEDGMENTS}

This work was supported in part by grant-in-aid for scientific research 
(Nos. 20115009 and 21740283)
from the Ministry of Education, Culture, Sports, Science and Technology (MEXT), Japan.

%% The Appendices part is started with the command \appendix;
%% appendix sections are then done as normal sections
\appendix

\section{Generating function for counting statistics}
\label{sec_appendix}

Here, we give a brief explanation for the counting statistics
for readers' convenience (For details, see \cite{Gopich2003,Gopich2005,Gopich2006}.)
In the framework of counting statistics, the quantity of interest is the number of target transitions.
It is needed to set multiple target transitions in the genetic switches,
and the genetic switches have two states, i.e., active and inactive states.
In the following explanations,
a simple setting, in which there is only one transition matrix and only one target transition,
will be discussed
because it is straightforward to apply the following simple discussions to the genetic switches.

Let $\{K_{nm}\}$ be a transition matrix.
We here derive the generating function for counting the number of events 
of a {\it specific} target transition $i_\mathrm{A} \to j_\mathrm{A}$.
Denote the probability, with which the system starts from state $m$ and finishes in state $n$
with $k$ transitions from $i_\mathrm{A}$ to $j_\mathrm{A}$ during time $t$, 
as $P_{nm}(k|t)$.
In order to calculate the probability $P_{nm}(k|t)$, we here define
a probability $G_{kl}'(t)$ with which the system evolves from state $l$ to state $k$, 
provided no $i_\mathrm{A} \to j_\mathrm{A}$ transitions occur during time $t$.
By using the probability $G_{kl}'(t)$,
the probability $P_{nm}(k|t)$ is calculated as
\begin{align}
&P_{nm}(k | t ) = \nonumber \\
&\quad G_{n j_\mathrm{A}}'(t) \ast 
\underbrace{K_{j_\mathrm{A} i_\mathrm{A}}(t) G_{i_\mathrm{A} j_\mathrm{A}}'(t) \ast \cdots 
\ast K_{j_\mathrm{A} i_\mathrm{A}}(t) G_{i_\mathrm{A} j_\mathrm{A}}' (t)
}_{k - 1} \nonumber \\
&\quad \ast K_{j_\mathrm{A} i_\mathrm{A}}(t) G_{i_\mathrm{A} m}' (t),
\end{align}
where $g_1(t) \ast g_2(t) \equiv \int_0^t g_1(t-t') g_2(t') d t'$ denotes the convolution.
This formulation means that an occurrence of the target transition $i_\mathrm{A} \to j_\mathrm{A}$
is sandwiched in between situations with no occurrence of the target transition,
and it is repeated $k$ times.

Next, we construct the generating function $\tilde{\phi}_{nm}(\chi,t)$
of the probability $P_{nm}(k|t)$:
\begin{align}
\tilde{\phi}_{nm}(\chi,t) = \sum_{k=0}^\infty \lambda^k P_{nm}(k|t).
\end{align}
That is, the generating function $\tilde{\phi}_{nm}(\lambda,t)$ 
gives the statistics of the number of transition $i_\mathrm{A} \to j_\mathrm{A}$ during time $t$
under the condition that the system starts from state $m$ and ends in state $n$.
The generating function $\tilde{\phi}_{nm}(\lambda,t)$ 
satisfies the following integral equation
\begin{align}
&\tilde{\phi}_{nm}(\lambda,t)  \nonumber \\
&= G_{nm}'(t) +
\int_0^t G_{n j_\mathrm{A}}' (t-t') \lambda K_{j_\mathrm{A} i_\mathrm{A}}(t')
\tilde{\phi}_{i_\mathrm{A} m} (\lambda,t') d t',
\end{align}
and obeys the following time-evolution equation
\begin{align}
&\frac{d}{d t} \tilde{\phi}_{nm} (\lambda,t) \nonumber \\
&\quad = \sum_i K_{ni}(t) G_{im}'(t) 
- \delta_{n, j_\mathrm{A}} K_{j_\mathrm{A} i_\mathrm{A}}(t) G_{i_\mathrm{A} m}'(t) \nonumber \\
& \qquad + \lambda G_{n j_\mathrm{A}}'(0) K_{j_\mathrm{A} i_\mathrm{A}}(t)  \tilde{\phi}_{i_\mathrm{A} m}(t)
\nonumber \\
& \qquad + \int_0^t \left( \frac{d}{d t} G_{n j_\mathrm{A}}'(t-t')\right)
\lambda K_{j_\mathrm{A} i_\mathrm{A}}(t') \tilde{\phi}_{i_\mathrm{A} m}(t') dt' \nonumber \\
&\quad =
\sum_i K_{ni}(t) \tilde{\phi}_{im}(\lambda,t) - \delta_{n, j_\mathrm{A}} (1-\lambda) 
K_{j_\mathrm{A} i_\mathrm{A}}(t) \tilde{\phi}_{i_\mathrm{A} m}(\lambda,t),
\label{eq_appendix_time_evolution_for_phi_nm}
\end{align}
where $\tilde{\phi}_{nm}(\lambda,0) = \delta_{n,m}$.
In order to show \eqref{eq_appendix_time_evolution_for_phi_nm},
we used the following two facts.
Firstly, 
the probability of no target transitions, $G_{nm}'(t)$, obeys
\begin{align}
\frac{d}{d t} G_{nm}'(t) = \sum_i K_{ni} (t) G_{im}'(t)
- \delta_{n, j_\mathrm{A}} K_{j_\mathrm{A} i_\mathrm{A}} (t) G_{i_\mathrm{A} m}'(t),
\end{align}
where $G_{nm}'(0) = \delta_{n,m}$.
Secondly, 
the derivative of the convolution is given by 
\begin{align}
&\frac{d}{d t} \int_0^t g_1(t-t') g_2(t') d t' \nonumber \\
&= g_1(0) g_2(t) + \int_0^t  \left( \frac{d}{d t} g_1(t - t') \right) g_2(t') d t'.
\end{align}

Using the generating function $\tilde{\phi}_{nm}(\lambda,t)$,
we construct restricted generating functions $\{\phi_n(\lambda,t)\}$ as follows:
\begin{align}
\phi_n(\lambda,t) = \sum_m \tilde{\phi}_{nm}(\lambda,t) p_m(0),
\label{appendix_eq_def_phi_n}
\end{align}
where $p_m(0)$ is a probability distribution at initial time $t=0$.
From \eqref{eq_appendix_time_evolution_for_phi_nm} 
and \eqref{appendix_eq_def_phi_n},
the restricted generating function satisfies
\begin{align}
&\frac{d}{d t} \phi_{n}(\lambda,t) \nonumber \\
&= \sum_{i} K_{ni}(t) \phi_{i}(\lambda,t) - \delta_{n,j_\mathrm{A}} (1-\lambda) 
K_{j_\mathrm{A}i_\mathrm{A}}(t)
\phi_{i_\mathrm{A}}(\lambda,t),
\end{align}
and these equations should be solved with initial conditions 
$\phi_n(\lambda,0) = \sum_m \tilde{\phi}_{nm} (\lambda,0) p_m(0)  = p_n(0)$.
The summation of $\{\phi_n(\lambda,t)\}$ for $n$ gives
the objective generating function for counting the number of events
of the specific target transition.

%In our case, there are infinite number of target transitions,
%because the number of proteins are not restricted to a finite number.
%we count the inactive $\to$ active transitions for all states
%Hence, the counting variable $\lambda$ should be multiplied 
%to all corresponding non-diagonal elements of the transition matrix,
%and finally we obtain Eqs.~\eqref{eq_cs_1_exact} and \eqref{eq_cs_2_exact}.

\end{document}